\documentclass[11pt]{article}
\usepackage{mcite, citesort}
\usepackage{amsmath,amssymb}
\usepackage{rotating}
\usepackage{caption}

\def\eo{\overset{_{\phantom{.}\circ}}{e}{}}

\def\go{\overset{_{\phantom{.}\circ}}{g}{}}
\def\Ko{\overset{_{\phantom{.}\circ}}{K}{}}
\def\To{\overset{_{\phantom{.}\circ}}{T}{}}
\def\Do{\overset{_{\phantom{.}\circ}}{D}{}}

\def\So{\overset{_{\phantom{.}\circ}}{S}{}}
\def\Ro{\overset{_{\phantom{.}\circ}}{R}{}}
\def\etao{\overset{_{\phantom{.}\circ}}{\eta}{}}
\newcommand{\fr}{\mathfrak{f}_{\scriptscriptstyle{F \hspace{-0.8mm} R}}}

\begin{document}

\begin{titlepage}

\hfill AEI-2013-114

\vspace{2cm}
\begin{center}

{{\LARGE  \bf 	Testing the non-linear flux ansatz for\\[3mm] maximal supergravity}} \\

\vskip 1.5cm {Hadi Godazgar, Mahdi Godazgar
and Hermann Nicolai}
\\
{\vskip 0.5cm
Max-Planck-Institut f\"{u}r Gravitationsphysik, \\
Albert-Einstein-Institut,\\
Am M\"{u}hlenberg 1, D-14476 Potsdam, Germany}
{\vskip 0.35cm
\texttt{Hadi.Godazgar@aei.mpg.de, Mahdi.Godazgar@aei.mpg.de,\\
Hermann.Nicolai@aei.mpg.de}}
\end{center}

\vskip 0.35cm

\begin{center}
\today
\end{center}

\noindent

\vskip 1cm

\begin{abstract}
\noindent We put to test the recently proposed non-linear flux ansatz for maximal 
supergravity in eleven dimensions, which gives the seven-dimensional flux 
in terms of the scalars and pseudoscalars of maximal $N=8$ supergravity, 
by considering a number of non-trivial solutions of gauged supergravity for which the higher dimensional solutions are known. These include the G$_2$ and SU(4)$^-$ invariant stationary points.  The examples considered constitute a very non-trivial check of the ansatz, which it passes with remarkable success.
\end{abstract}

\end{titlepage}

\section{Introduction}

Recently \cite{dWN13}, a simple non-linear flux ansatz giving the seven-dimensional 
components of the 3-form potential of eleven-dimensional supergravity \cite{CJS} in terms of the scalars and pseudoscalars of maximal (gauged) $N=8$ supergravity \cite{dWNn8} has been proposed.  This result arose from an attempt to understand the embedding of a recently discovered continuous family of inequivalent maximal ($N=8$) gauged supergravities in four dimensions \cite{DIT}.
The emergence of this new family of theories follows from the electric-magnetic duality
of the ungauged $N=8$ theory \cite{cremmerjulia}, and can thus be understood in terms of the freedom to rotate between how one chooses to define electric and magnetic vector fields \cite{GZ}.  The inequivalence of the resulting theories is confined to the gauged theory, because in the ungauged theory electric-magnetic duality renders all such theories equivalent.  From an eleven-dimensional perspective, the electric vector fields arise from the off-diagonal elfbein components (graviphoton), while the magnetic vector fields emerge from particular components of the 3-form potential.

A standard method by which new theories are obtained in supergravity is by reducing a higher dimensional theory on some group manifold or coset space.  The problem of determining whether a coset space reduction is consistent, in the sense that every solution of the lower dimensional theory can be uplifted to a solution of the higher dimensional theory is a subtle one.  In fact, the expectation is that such reductions are, in general, inconsistent \cite{garydeserfest}.  A notable exception to this expectation is the consistency of the seven-sphere reduction of eleven-dimensional supergravity \cite{dWNconsist87, NP}.  Central to this result is a local SU(8) invariant reformulation of the eleven-dimensional theory \cite{dWNsu8}, which in the reduction on
a seven-torus $T^7$ immediately reduces to the E$_7$ invariant theory of 
Cremmer and Julia \cite{cremmerjulia}, without the need to dualise tensors to scalars.
 This reformulation necessitates the introduction 
of new SU(8) covariant objects in eleven dimensions.  The most significant such object is the generalised vielbein, which arises from the study of the supersymmetry transformation of the graviphoton and replaces the siebenbein in the reformulated theory.  The intimate connection between the electric vector fields of the four dimensional theory and the graviphoton 
leads naturally to the non-linear metric ansatz \cite{dWNW}

\begin{equation}
\Delta^{-1} g^{mn}(x,y) = \frac{1}{8} K^{m}{}^{IJ}(y) K^{n}{}^{KL}(y) 
\Big[ (u^{ij}{}_{IJ} + v^{ijIJ}) (u_{ij}{}^{KL} + v_{ijKL}) \Big](x),
\label{metansatz}
\end{equation}
whereby the seven-dimensional metric $g_{mn}$ is given terms of the scalar and 
pseudoscalar fields of $N=8$ supergravity, via the E$_{7(7)}$ matrix components
$u(x)$ and $v(x)$ (see \eqref{uv} below), and of the Killing vectors 
$K^m(y)$ on the seven-sphere $S^7$
(where the eleven-dimensional coordinates are split as $z^M = (x^\mu, y^m)$).
We note that the above formula has been subjected to numerous tests and also 
proven its usefulness in other contexts, such as the AdS/CFT correspondence \cite{dWNW, memflow, ahn01, ahn02, Gowdigere, popedielec, bobev10, bobev11}.  The proof of consistency in \cite{dWNconsist87} also furnishes a formula for the 
4-form field strength, modulo a subtlety that is resolved in \cite{NP}.  However, this 
formula appears to be  too cumbersome for practical applications.

The remarkable result of \cite{dWN13} is that there exists an object analogous to 
the generalised vielbein that arises from the supersymmetry transformation of the components of the 3-form potential from which the magnetic vector fields of the four-dimensional theory arise.  This new generalised vielbein now replaces the components of the 3-form potential $A_{mnp}$ along the seven directions in the local SU(8) invariant reformulation of the eleven-dimensional theory.  Furthermore, it leads to a non-linear flux ansatz, which complements the non-linear metric ansatz above \cite{dWN13}:
\begin{align}
\sqrt{2} \,K^{p}{}^{IJ}(y) K^{q}{}^{KL}(y) &\Big[\left( u^{ij}{}_{IJ} + v^{ij IJ} \right) \left( u_{ij}{}^{KL} + v_{ij KL} \right)\Big](x) A_{mnp}(x,y) \notag \\[7pt]
 &=- i K_{mn}{}^{IJ}(y) K^{q}{}^{KL}(y) \Big[\left( u^{ij}{}_{IJ} - v^{ij IJ} \right) \left( u_{ij}{}^{KL} + v_{ij KL} \right)\Big](x). \label{potans}
\end{align}
While this non-linear flux ansatz takes a surprisingly simple form, it is not a formula that can be found by asserting consistency with previous results. This is a major difference with the corresponding result for the AdS$_7 \times S^4$ compactification of maximal supergravity, 
where the non-linear ans\"atze can be directly substituted into the higher-dimensional field
equations \cite{S4consislet, *S4consis}; such a direct substitution is not possible 
for the AdS$_4 \times S^7$ compactification. Let us also mention that there exist 
partial results and uplift formulae for truncated versions of the maximal theory where the scalar sector is much simpler (see for example \cite{Cvetic:2000dm, Cvetic:1999au, Cvetic:2000tb, Gauntlett:2009zw, Bah:2010yt, OColgain:2011ng, Cassani:2012pj, Borghese:2012zs} and references therein).  However, the formula above cannot be guessed from these.  Its derivation is critically dependent on an analysis of the local SU(8) manifest formulation of eleven-dimensional supergravity.  Last but not least we wish to point out that in comparison with other theories, 
$N=8$ supergravity is distinguished by an astonishingly rich variety of stationary points \cite{Fischbachernew, *Fischbacherency}~\footnote{Whereas, for instance, maximal gauged supergravity in seven dimensions has only 
one non-trivial stationary point besides the trivial vacuum \cite{s4critical}.} 
that can now be explored by means of the new formula.
     
The non-linear flux ansatz given in equation (5.11) of \cite{dWN13} differs from the above 
expression by the factor of $\sqrt{2}$ on the left hand side.  In fact, as already pointed 
out there, this overall factor can so far not be determined from intrinsically
Kaluza-Klein theoretic considerations matching the eleven-dimensional 3-form potential 
with the four-dimensional gauge field. This is in marked contrast to the graviphoton, for which general 
Kaluza-Klein theory gives its precise relation to the four dimensional gauge field, 
by matching the non-abelian interaction with the commutator of two Killing vector fields. 
However, as we will show here, this factor is universally and unambiguously 
the same for all solutions, and  does follow by explicitly computing the 3-form 
potential using the above ansatz for solutions of gauged supergravity for which 
the higher dimensional uplift solution is known (the factor $\sqrt{2}$ is most
easily checked  for the Englert solution \cite{englert}).

The aim of this paper, then, is to test the non-linear flux ansatz for a number of solutions of gauged supergravity.  In order to make the paper self-contained we begin, in section \ref{sec:prelim}, by describing the main conventions and definitions that are required, summarising 
various known results. In addition, in appendix \ref{app:gamma}, we list some important 
$\Gamma$-matrix identities, most of which already appear in the appendices 
of \cite{cremmerjulia} and \cite{dWNsu8}, while some are new.

In section \ref{sec:g2}, we begin by describing the eleven-dimensional G$_2$ invariant solution of \cite{dWNW}.  Then, we consider the G$_2$ invariant stationary point of $N=8$ supergravity \cite{Warner83, Warner84} that uplifts to the aforementioned solution, rederiving the E$_7$ matrix components $u$ and $v$ that are parametrised by the scalars and pseudoscalars.  These components are essential inputs in the non-linear ans\"atze.  We calculate the 3-form potential using the non-linear flux ansatz, equation \eqref{potans}, verifying its total antisymmetry as is expected from the general argument in \cite{dWN13}.  The field strength of this potential is then derived for the G$_2$ family of solutions.  Substituting the G$_2$ stationary point values yields the flux of the G$_2$ invariant solution of the eleven-dimensional theory with precise agreement.

As our next test, we consider the SU(4)$^-$ invariant solution of \cite{PW} in section \ref{sec:su4}. We rewrite this solution of eleven-dimensional supergravity in terms of geometric quantities defined on the seven-sphere.  As in section \ref{sec:g2}, we derive the 3-form potential 
using the non-linear flux ansatz and confirm that at the stationary point \cite{Warner83, Warner84} the associated field strength matches precisely with that of the eleven-dimensional solution.

Furthermore, in appendix \ref{app:so7}, we give the metric and the flux calculated from the non-linear ans\"atze with the scalars of the SO(7)$^{\pm}$ invariant family of maximal gauged 
supergravity \cite{Warner83, Warner84}.  These examples  are simple enough for the 
reader to immediately match with the known SO(7)$^+$ \cite{dWNso7soln} and SO(7)$^-$ \cite{englert} solutions of eleven-dimensional supergravity, and are thus included
mainly for the reader's convenience.

\section{Preliminaries} \label{sec:prelim}

In this paper, we follow the conventions of reference \cite{dWNsu8}.  The bosonic field 
equations of eleven-dimensional supergravity \cite{CJS} read\footnote{Note that for consistency 
with \cite{dWNW}, we use a negative curvature convention, i.e. $[D_M,D_N]V^P=-{R^{P}}_{QMN}V^{Q}$.  Hence, the scalar curvature of a sphere is negative.} 
\begin{gather}
 R_{MN} = \textstyle{\frac{1}{72}} g_{MN} F^2_{PQRS} - \textstyle{\frac{1}{6}} F_{MPQR} {F_N}^{PQR}, \label{11dEinstein} \\[7pt]
 E^{-1} \partial_{M} (E F^{MNPQ}) = \textstyle{\frac{\sqrt{2}}{1152}} i \eta^{NPQR_1\ldots R_4 S_1 \ldots S_4} F_{R_1\ldots R_4} F_{S_1\ldots S_4}, \label{11dMaxwell}
\end{gather}
where $E$ is the determinant of the elfbein ${E_M}^A$.  We note that solutions to
these combined equations are only determined up to an overall constant scaling
\begin{equation}\label{scaling}
g_{MN} \;\; \rightarrow \;\; \lambda \, g_{MN}, \qquad
F_{MNPQ} \;\; \rightarrow \;\; \lambda^{3/2} \, F_{MNPQ}.
\end{equation}
Such a rescaling must be taken into account when comparing the various solutions 
given in the literature with the ones constructed from the non-linear 
ans\"atze \eqref{metansatz} and \eqref{potans}. We emphasise that the normalisation
of all solutions is thus completely fixed by \eqref{metansatz} and \eqref{potans}, once 
the trivial vacuum solution has been specified.

We are interested in solutions of the above equations that are obtained via a compactification to a four-dimensional maximally symmetric  spacetime.  The most general ansatz for the elfbein that is consistent with this requirement is of the warped form\footnote{In general, of course, 
  ${e_m}^a$ can also have $x$ dependence. But here we are considering compactifications.} 
\begin{equation} \label{elfbein}
 {E_M}^A(x,y) = \begin{pmatrix}
                 \Delta^{-1/2}(y) {\eo_{\mu}}{}^{\alpha}(x) & 0 \\ 0 & {e_m}^a(y)
                \end{pmatrix},
\end{equation}
where $x^\mu$ are coordinates on the four-dimensional spacetime and $y^m$ are coordinates on the compact seven-dimensional space; ${\eo}_{\mu}{}^{\alpha}(x)$ is the vierbein of the maximally symmetric four-dimensional spacetime and ${e_m}^a(y)$ is the siebenbein of the compact space.  In particular, we assume the siebenbein to be that of a deformed round seven-sphere with the deformation parametrised by a matrix ${S_a}^b(y)$
\begin{equation}\label{eS}
 {e_m}^a(y)={\eo_m}{}^b (y) \, {S_b}^a(y),
\end{equation}
where ${\eo_m}{}^a (y)$ is the siebenbein on the round seven-sphere with 
inverse radius $m_7$, and where
\begin{equation}
 \Delta(y) \, \equiv \,\textup{det} \, {S_a}^b(y).
\end{equation}
The presence of the warp factor in \eqref{elfbein} is required by consistency with the supersymmetry transformation rules of the fields that would correspond with those 
of the maximal theory upon reduction to four dimensions \cite{dWNso7soln}.

The eight Killing spinors of $S^7,$ $\eta^{I}$ satisfy 
\begin{equation} \label{spinoreqn}
 \big(\Do_{m} + \frac{1}{2} i \, m_{7} \eo_{m}{}^{a} \Gamma_{a}\big) \eta^{I} = 0,
\end{equation}
where $\Do_{m}$ is the covariant derivative on the round seven-sphere and the $\Gamma^{a}$-matrices are flat, Euclidean, seven-dimensional and purely imaginary.  In a Majorana representation of the Clifford algebra in Euclidean seven dimensions, the charge conjugation matrix that is used to define spinor conjugates, or raise and lower spinor indices, can be chosen to be the identity matrix. Here we make such a choice. Furthermore, it is useful to choose Killing spinors that are orthonormal 
\begin{equation}
 \bar{\eta}^{I} \eta^{J} = \delta^{IJ}, \qquad  \eta^{I} \bar{\eta}^{I}  =  {\bf 1} ,
\label{KSorthnorm}
\end{equation}
where on the right hand side of the second equation, $\bf 1$ denotes the identity 
matrix with spinor indices. 

These spinors can be used to define a set of vectors and 2-forms as follows:
\begin{align}
 K_{a}{}^{IJ} &= i \bar{\eta}^{I} \Gamma^{a} \eta^{J}, \qquad \ \  K_{ab}{}^{IJ} = \bar{\eta}^{I} \Gamma^{ab} \eta^{J}, \label{Kadef} \\
 K_{m}{}^{IJ} &= \eo_m{}^a K_a{}^{IJ}, \qquad K_{mn}{}^{IJ} = \eo_m{}^a \eo_n{}^b K_{ab}{}^{IJ}.
\end{align}
In the following we will adopt the rule that the curved indices on Killing vectors and their
derivatives are always lowered and raised {\em with the round seven-sphere metric
$\go_{mn}$ and its inverse.} It is now straightforward to show that
\begin{align}
 K_{ab}{}^{IJ} K_{c}{}^{IJ} = 0, \qquad K^{a}{}^{IJ} K_{b}{}^{IJ} = 8\, \delta^{a}_{b}, \qquad K^{ab}{}^{IJ} K_{cd}{}^{IJ} = 16 \, \delta^{ab}_{cd}. \label{Kcontract}
\end{align}

Assuming the four-dimensional spacetime to be maximally symmetric implies that the only non-zero components of the field strength $F_{MNPQ}$ are $F_{\mu \nu \rho \sigma}$ and $F_{mnpq}$.  Following \cite{freundrubin}, we parametrise $F_{\mu \nu \rho \sigma}$ as follows
\begin{equation} \label{4dFlux}
 F_{\mu \nu \rho \sigma}= i \fr \eta_{\mu \nu \rho \sigma},
\end{equation}
where $\eta_{\mu \nu \rho \sigma}$ is the alternating tensor in four dimensions.\footnote{Note that the conventions used in this paper are such that $\eta_{\mu \nu \rho \sigma} \eta^{\mu \nu \rho \sigma} = + 4!$.}  The Bianchi identities imply that the Freund-Rubin parameter $\fr$ is a constant.  Beware that switching to flat indices introduces $y$-dependence
\begin{equation}
 F_{\alpha \beta \gamma \delta}= i \fr \Delta^{2} \, \eta_{\alpha \beta \gamma \delta}.
\end{equation}

Given an elfbein of the form given in equation \eqref{elfbein} and using equation \eqref{4dFlux}, it is fairly straightforward to show that the eleven-dimensional equations \eqref{11dEinstein} and \eqref{11dMaxwell} reduce to \cite{dWNW}
\begin{gather}
{R_{\mu}}^{\nu}=\Big(\textstyle{\frac{2}{3}} \fr^2 \Delta^4 + \textstyle{\frac{1}{72}} F_{mnpq}^2 \Big) \delta_{\mu}^{\nu}, \label{Einseqn1} \\[6pt]
{R_{m}}^{n}= -\textstyle{\frac{1}{6}} F_{mpqr} F^{npqr} + \Big(\textstyle{\frac{1}{72}} F_{mnpq}^2 - \textstyle{\frac{1}{3}} \fr^2 \Delta^4 \Big) \delta_{m}^{n}, \label{Einseqn2} \\[6pt]
\Do_q \big(\Delta^{-1} F^{mnpq}\big)=\textstyle{\frac{1}{24}\sqrt{2}} \fr \etao^{mnpqrst}F_{qrst} \label{maxwell},
\end{gather}
where seven-dimensional indices $m, n, p, \ldots$ are raised (lowered) with 
$$g^{mn}={e_a}^m {e_b}^n \delta^{ab} \ \ (g_{mn}={e_m}^a {e_n}^b \delta_{ab}),$$
except in cases where the object is denoted with a circle ${}^\circ$ on top, in which case indices are raised (lowered) with $\go{}^{mn}$ ($\go_{mn}$) analogously defined.  Hence, $\etao_{mnpqrst}$ is the alternating tensor corresponding to the round seven-sphere metric $\go_{mn}$ and its indices are raised with $\go^{mn}$.

As is well-known, the four-dimensional spacetime must be AdS$_4$.  We choose to parametrise its radius such that
\begin{equation}
 R_{\mu \nu}=3 m_4^2 g_{\mu \nu}.
\end{equation}
Furthermore, for an $S^7$ of inverse radius $m_7$,
\begin{equation}
 \Ro_{mn}=-6 m_7^2 \go_{mn}. \label{riccis7}
\end{equation}
Thus, in our conventions, the $S^7$ compactification \cite{duffpope} is given by 
\begin{equation}
 m_4=2 m_7, \qquad \fr=\pm 3\sqrt{2} m_7.
\end{equation}
We repeat that the normalisation of {\em all} solutions away from the trivial AdS$_4$ vacuum 
is fixed by the non-linear ans\"atze. Thus, they are all expressed in terms of a single dimensionful parameter $m_7$.

\section{The G$_2$ invariant solution} \label{sec:g2}

\subsection{The G$_2$ invariant solution of eleven-dimensional supergravity}

In order to write out the G$_2$ invariant solution, we must first define the geometrical quantities,
respectively preserving the SO(7)$^+$ and SO(7)$^-$ subgroups of SO(8) whose
common subgroup is G$_2 =$ SO(7)$^+ \cap$ SO(7)$^-$. These are given in terms of
the following self-dual $C_{+}^{IJKL}$ and antiself-dual $C_{-}^{IJKL}$ SO(8) tensors, respectively, which satisfy the identities \cite{dWNso7soln,parallel}
\begin{align}
 C_{+}^{IJMN} C_{+}^{MNKL} &= 12 \delta^{IJ}_{KL} + 4 C_{+}^{IJKL}, \label{C+C+} \\
 C_{-}^{IJMN} C_{-}^{MNKL} &= 12 \delta^{IJ}_{KL} - 4 C_{-}^{IJKL} \label{C-C-}.
\end{align}
These tensors will also appear below in the parametrisation of the scalar and pseudoscalar 
expectations in $N=8$ supergravity.

The self-dual tensor $C_{+}$ can be used to define SO(7)$^{+}$ invariant 
quantities \cite{dWNso7soln} 
\begin{align}
 \xi_{a} &= \frac{1}{16} C_{+}^{IJKL} K_{ab}{}^{IJ} K_{b}{}^{KL}, \label{defxia} \\
 \xi_{ab} &= - \frac{1}{16} C_{+}^{IJKL} K_{a}{}^{IJ} K_{b}{}^{KL}, \label{defxiab} \\[5pt]
 \xi &= \delta^{ab} \xi_{ab}. \label{defxi} 
\end{align}
These quantities satisfy the non-trivial identities \cite{dWNso7soln} 
\begin{align}
 \xi_{a} \xi_{a} &= (21 + \xi)(3 - \xi), \\
\xi_{ab} &= \frac{1}{6} (3 + \xi) \delta_{ab} - \frac{1}{6(3-\xi)} \xi_{a} \xi_{b}, \\
\Do_{c} \xi_{ab} &= \frac{1}{3} m_7 \left(\delta_{ab} - \xi_{(a} \delta_{b)c} \right), \\
\Do_{a} \xi &= 2 m_7 \xi_{a}, \\ 
\Do_{a} \xi_{b} &= m_7 (3 - \xi) \delta_{ab} - \frac{m_7}{3-\xi} \xi_{a} \xi_{b}.   
\end{align}
Hence, the variable $\xi$ lies in the range $- 21 < \xi < 3$, with the endpoints corresponding to the north 
and south poles of the seven sphere. Alternatively, in terms of the unit vector 
\begin{equation}
 \hat{\xi}_{a} = \frac{1}{\sqrt{(21 + \xi)(3 - \xi)}} \xi_{a},
\end{equation}
 the last two equations become \cite{dWNso7soln} 
\begin{align}
\Do_{a} \xi &= 2 m_7 \sqrt{(21 + \xi)(3 - \xi)} \hat{\xi}_{a}, \label{Dxi} \\ 
\Do_{a} \hat{\xi}_{b} &= m_7 \sqrt{\frac{3 - \xi}{21+ \xi}} \left( \delta_{ab} - \hat{\xi}_{a} \hat{\xi}_{b} \right). \label{Dxia}   
\end{align}

The antiself-dual $C_{-}^{IJKL}$ can similarly be used to define the SO(7)$^{-}$ 
invariant tensor ({\it alias} the `parallelising torsion' on $S^7$)
\begin{equation}
 S_{abc} = \frac{1}{16} C_{-}^{IJKL} K^{IJ}_{[ab} K^{KL}_{c]}, \label{Sdef}
\end{equation}
 which satisfies the relations
\begin{align}
 \Do_{a} S_{bcd} = \frac{1}{6} m_{7} \epsilon_{abcdefg} S^{efg}, \label{DS} \\
 S^{[abc} S^{d]ef} = \frac{1}{4} \epsilon^{abcd[e}{}_{gh} S^{f]gh}, \label{SSanti1} \\
 S^{a[bc} S^{de]f} = \frac{1}{6} \epsilon^{bcde(a}{}_{gh} S^{f)gh}, \label{SSanti2} \\
 S^{abe} S_{cde} = 2 \delta^{ab}_{cd} + \frac{1}{6} \epsilon^{ab}{}_{cdefg} S^{efg} \label{Ssquared}.
\end{align}
These relations have been derived in \cite{parallel, rel411}.  There is a potential ambiguity in the sign of terms with $S$ on the right hand side of the equations above, which is fixed by requiring that $C_{-}$ is antiself-dual and satisfies equation \eqref{C-C-} (see equations (3.6)--(3.17) of \cite{parallel}).  Equation \eqref{DS} is derived using the $\Gamma$-matrix 
identity \eqref{gammaabcdual2} and equation \eqref{spinoreqn}.

The relations \eqref{defxia}, \eqref{defxiab} and \eqref{Sdef} can be inverted to give the SO(8) tensors in terms of the SO(7)$^{+}$ and SO(7)$^{-}$ geometric quantities \cite{rel411,dWNso7soln,NP}
\begin{align}
 C_{+}^{IJKL} &= - \frac{1}{12} (9 + \xi) K_{a}{}^{[IJ} K_{a}{}^{KL]} + \frac{1}{4}(21+\xi) \hat{\xi}^{a} \hat{\xi}^{b} K_{a}{}^{[IJ} K_{b}{}^{KL]} \notag \\
& \hspace{55mm} + \frac{1}{12} \sqrt{(21 + \xi)(3 - \xi)} \hat{\xi}^{a} K_{ab}{}^{[IJ} K_{b}{}^{KL]}, \label{C+}\\
 C_{-}^{IJKL} &= \frac{1}{2} S_{abc} K_{ab}{}^{[IJ} K_{c}{}^{KL]}. \label{C-}
\end{align}
In appendix \ref{app:crelns}, we explain that $C_{+}, iC_{-}$ together with their symmetrised product
$iD_{+}$ generate the SU(1,1) algebra in E$_7$ \cite{dWNW}, which commutes with G$_2$.
This fact can be used to used derive the relations listed in equations \eqref{C+D+}--\eqref{D-D-}. 

In terms of the SO(7)$^{\pm}$ invariant tensors defined above, the G$_2$ invariant solution of
eleven-dimensional supergravity is given by the following expressions. In the 
uncompactified dimensions, it is the usual AdS$_4$ metric, while the  
metric in the internal seven-dimensional space is given by \cite{dWNW}
\begin{equation} \label{11dmetgam}
 g_{mn} = 6^{2/3} \gamma^{-1/9} (15-\xi)^{-1/3} \left\lbrace \left(\go_{mn} - \hat{\xi}_m \hat{\xi}_n\right) + \textstyle{\frac{1}{36}}(15 - \xi)\hat{\xi}_m \hat{\xi}_n \right\rbrace,
\end{equation}
where $\gamma$ is an arbitrary positive constant and the index on $\hat{\xi}_m$ is raised with metric $\go^{mn}$.  The determinant of this metric is
\begin{equation}
 \textup{det}(g_{mn})=\Delta^2 \ \textup{det}(\go_{mn}),
\end{equation}
where
\begin{equation} \label{def:Delta}
 \Delta = 6^{4/3} \gamma^{-7/18} (15-\xi)^{-2/3}.
\end{equation}
The internal flux (4-form field strength) is
\begin{align}
 F_{mnpq} =\frac{4 \sqrt{6/5}}{15-\xi} \gamma^{-1/6} m_7 &\left\lbrace \etao_{mnpqrst} \So^{rst} - \frac{(21+\xi)(\xi-27 \pm 12 \sqrt{3})}{12(15-\xi)}  \hat{\xi}_{[m} \etao_{npq]rstu} \hat{\xi}^r \So^{stu} \right. \notag \\[6pt]
 &\hspace{25mm} +\left. \sqrt{(21+\xi)(3-\xi)} \frac{(\xi-51 \pm 12 \sqrt{3})}{2(15-\xi)} \So_{[mnp} \hat{\xi}_{q]} \right\rbrace. \label{11dflux}
\end{align}
The $\pm$ ambiguity in the expression above arises from the arbitrariness in the sign of the Freund-Rubin parameter $\fr$ \cite{dWNW}. As shown there, this solution has $N=1$ residual supersymmetry.

The solution given above solves the Einstein equations for any value of the constant $\gamma$ (see equation \eqref{scaling}). However, the non-linear metric ansatz gives the solution with a particular value for $\gamma$. In anticipation of this fact, and for ease of comparison later, we choose 
\begin{equation}
 \gamma^{-1/3}=\frac{5}{6\sqrt{3}}.
\end{equation}
Hence,
\begin{equation} \label{11dmet}
 g_{mn} = 3^{1/6} 10^{1/3} (15-\xi)^{-1/3} \left\lbrace \left(\go_{mn} - \hat{\xi}_m \hat{\xi}_n\right) + \textstyle{\frac{1}{36}}(15 - \xi)\hat{\xi}_m \hat{\xi}_n \right\rbrace
\end{equation}
and
\begin{align}
 F_{mnpq} =\frac{4\times 3^{-1/4}}{15-\xi} m_7 &\left\lbrace \etao_{mnpqrst} \So^{rst} - \frac{(21+\xi)(\xi-27 \pm 12 \sqrt{3})}{12(15-\xi)}  \hat{\xi}_{[m} \etao_{npq]rstu} \hat{\xi}^r \So^{stu} \right. \notag \\[6pt]
 &\hspace{25mm} +\left. \sqrt{(21+\xi)(3-\xi)} \frac{(\xi-51 \pm 12 \sqrt{3})}{2(15-\xi)} \So_{[mnp} \hat{\xi}_{q]} \right\rbrace.  \label{g2flux}
\end{align}

\subsection{The G$_2$ invariant stationary point of gauged supergravity}
\label{sec:G2gauged}

The 70 scalars and pseudoscalars of the $N=8$ supergravity theory that parametrise 
an element of the coset space E$_7$/SU(8) can be described by an element in the 
fundamental representation of E$_7$ as follows \cite{cremmerjulia}:
\begin{equation} \label{uv}
 \mathcal{V}=\begin{pmatrix} 
              {u_{ij}}^{IJ} & v_{ijIJ} \\
              v^{ijIJ} & {u^{ij}}_{IJ}
             \end{pmatrix}.
\end{equation}
Note that complex conjugation is represented by a respective lowering/raising of indices.

Using an SU(8) transformation, the E$_7$ matrix $\mathcal V$ can be brought into a symmetric gauge of the form
\begin{equation}
 \mathcal{V}=\text{exp} \, \Phi\equiv \text{exp}\begin{pmatrix}
              0 & \phi_{IJKL} \\
              \phi^{IJKL} & 0
             \end{pmatrix}.
\end{equation}
Once this gauge is fixed, the distinction between $i,j, \ldots$ and $I,J, \ldots$ indices may be safely ignored, as we shall do so hereafter. For a G$_2$ invariant configuration, the most general vacuum expectation value that $\phi_{IJKL}$ can take may be parametrised as follows \cite{Warner83, Warner84}:
\begin{equation}
 \phi_{IJKL} \equiv 
 \phi_{IJKL}(\lambda,\alpha)=\textstyle{\frac{1}{2}}\lambda\big(C_{+}^{IJKL} \cos \alpha + 
 i {C_{-}^{IJKL}} \sin \alpha \big), \label{phi}
\end{equation}
where $\lambda$ and $\alpha$ take a particular value for each stationary point consistent with this configuration.  The self-dual $C_{+}^{IJKL}$, antiself-dual $C_{-}^{IJKL}$ and 
\begin{equation}
 D_{\pm}^{IJKL} = \frac{1}{2} \left( C_{+}^{IJMN} C_{-}^{MNKL} \pm C_{-}^{IJMN} C_{+}^{MNKL} \right) \label{Dpmdef},
\end{equation}
form a basis of G$_2$ invariant objects in E$_7.$ In the remainder we will often
keep the index structure implicit for brevity; so
\begin{equation}\label{AB}
A\cdot B \equiv \big( A\cdot B \big)^{IJ\, KL} \equiv A^{IJMN} B^{MNKL}.
\end{equation}
Given $\phi$ of the form above, the components of the E$_7$/SU(8) coset elements $u^{IJ}{}_{KL}$ and $v^{IJKL}$ can be written in terms of the G$_2$ invariant $C_{\pm}$ and $D_{\pm}.$  Given the structure of the matrix $\Phi$, it is not too difficult to see that
\begin{align} \label{Vphiphistar} \renewcommand{\arraystretch}{2.5}
 \mathcal{V} &= \sum_{n=0}^{\infty} \frac{1}{n!} \Phi^n 
 = \begin{pmatrix} \renewcommand\arraystretch{4}
              \displaystyle{\sum_{n=0}^{\infty}} \frac{(\lambda/2)^{2n}}{(2n)!} (\varphi \varphi^*)^n & \displaystyle{\sum_{n=0}^{\infty}} \frac{(\lambda/2)^{2n+1}}{(2n+1)!} \varphi (\varphi^* \varphi)^n \\
              \displaystyle{\sum_{n=0}^{\infty}} \frac{(\lambda/2)^{2n+1}}{(2n+1)!} \varphi^* (\varphi \varphi^*)^n & \displaystyle{\sum_{n=0}^{\infty}} \frac{(\lambda/2)^{2n}}{(2n)!} (\varphi^* \varphi)^n
	     \end{pmatrix},
\end{align}
where
\begin{equation} \label{varphi}
 \varphi=\cos \alpha \, C_+  + i\sin \alpha \, C_-, \quad \varphi^*=\cos \alpha \, C_+  - i\sin \alpha \, C_-.
\end{equation}
Of course, the order in which the tensors appear is now important and indicative of the index structure of the terms.  For example,  $C_+ C_{-} = D_+ + D_-$, while $C_{-} C_+ = D_+ - D_-$.  The matrix above is clearly compatible with the structure of the matrix given in the defining equation \eqref{uv}.

Consider
\begin{equation}
 \varphi \varphi^*= \big(\cos \alpha \, C_+  + i\sin \alpha \, C_- \big)
 \big(\cos \alpha \, C_+  - i\sin \alpha \, C_- \big).
\end{equation}
Using equations \eqref{C+C+} and \eqref{C-C-}, we find that
\begin{equation} \label{phiphistar}
 \varphi \varphi^* = 12 + 4\tilde{\Theta},
\end{equation}
where we have omitted a $\delta$ symbol in the first term above for brevity and
\begin{equation} \label{tiltheta}
 \tilde{\Theta} = \cos^2 \alpha \, C_+  -\sin^2 \alpha \, C_-  - \textstyle{\frac{1}{4}} i \sin 2\alpha \, D_-.
\end{equation}
We notice that $\tilde{\Theta}$ has the rather convenient property that
\begin{equation}
 \tilde{\Theta}^2 = 12 + 4  \tilde{\Theta}.
\end{equation}
One can simply verify the above equation using equations \eqref{C+D+}--\eqref{D-D-}. Now, define a new quantity
\begin{equation} \label{deftheta}
  \Theta = \textstyle{\frac{1}{8}} (\tilde{\Theta} +2),
\end{equation}
which has been chosen so that
\begin{equation} \label{powerprop}
 \Theta^2=\Theta.
\end{equation}
From equation \eqref{phiphistar}, we have
\begin{equation} \label{phiphistartheta}
 \varphi \varphi^* = 12 + 4 \tilde{\Theta} = 4 + 32 \Theta.
\end{equation}
Comparing the components of equations \eqref{uv} and \eqref{Vphiphistar} gives that
\begin{equation}
  {u_{IJ}}^{KL} = \displaystyle{\sum_{n=0}^{\infty}} \frac{(\lambda/2)^{2n}}{(2n)!} (\varphi \varphi^*)^n
\end{equation}
and
\begin{equation} \label{vphi}
  v^{IJKL} =   \displaystyle{\sum_{n=0}^{\infty}} \frac{(\lambda/2)^{2n+1}}{(2n+1)!} \varphi^* (\varphi \varphi^*)^n,
\end{equation}
where, as before, indices have been suppressed on the right hand side of the above equations.

First, consider ${u_{IJ}}^{KL}$.
\begin{align}
 {u_{IJ}}^{KL} &= \displaystyle{\sum_{n=0}^{\infty}} \frac{(\lambda/2)^{2n}}{(2n)!} (\varphi \varphi^*)^n \notag \\
	       &= \displaystyle{\sum_{n=0}^{\infty}} \frac{(\lambda/2)^{2n}}{(2n)!} (4 + 32 \Theta)^n \notag \\
	       &= \displaystyle{\sum_{n=0}^{\infty}} \frac{(\lambda/2)^{2n}}{(2n)!} \displaystyle{\sum_{p=0}^{n}} 
	          \begin{pmatrix} n \\ p \end{pmatrix} 4^{n-p} (32\Theta)^{p}, \notag
\end{align}	
where we have used equation \eqref{phiphistartheta} in the second equality and applied the binomial theorem in the third equality.  Using the property satisfied by $\Theta$, equation \eqref{powerprop}, the previous expression can be rewritten as follows:
\begin{align}
{u_{IJ}}^{KL}  &= \displaystyle{\sum_{n=0}^{\infty}} \frac{(\lambda/2)^{2n}}{(2n)!} \left\lbrace 4^n + \displaystyle{\sum_{p=1}^{n}} 
	          \begin{pmatrix} n \\ p \end{pmatrix} 4^{n-p} (32)^p \Theta \right\rbrace \notag \\
	       &= \displaystyle{\sum_{n=0}^{\infty}} \frac{(\lambda/2)^{2n}}{(2n)!} \left\lbrace 4^n + \left[ \displaystyle{\sum_{p=0}^{n}} 
	          \begin{pmatrix} n \\ p \end{pmatrix} 4^{n-p} (32)^p - 4^n \right] \Theta \right\rbrace 
	          \notag \\
	       &= \displaystyle{\sum_{n=0}^{\infty}} \frac{\lambda^{2n}}{(2n)!} 
	       \Big\lbrace 1 + \left[ 3^{2n} - 1 \right] \Theta \Big\rbrace. \notag 
\end{align}
Identifying the above expressions as the Taylor expansions of the $\cosh$ function simplifies the expression to
\begin{align}
{u_{IJ}}^{KL}  &= \cosh \lambda + (\cosh 3\lambda - \cosh \lambda) \Theta \notag \\
	       &= \cosh \lambda + \textstyle{\frac{1}{8}} (\cosh 3\lambda - \cosh \lambda) (\tilde{\Theta} + 2) \notag \\
	       &= \cosh^3 \lambda + \textstyle{\frac{1}{2}} \cosh \lambda \sinh^2 \lambda \, \tilde{\Theta}, \notag
\end{align}
where we have used equation \eqref{deftheta} in the second equality and well-known multiple angle identities for hyperbolic functions in the final equality.  Defining
\begin{equation} \label{defpq}
 p= \cosh \lambda, \quad q= \sinh \lambda
\end{equation}
and substituting for $\tilde{\Theta}$ using equation \eqref{tiltheta} gives
\begin{equation}
 {u_{IJ}}^{KL} (\lambda,\alpha) \,=\, p^3 \delta_{IJ}^{KL} + \textstyle{\frac{1}{2}} p q^2 \cos^2 \alpha \, C_+^{IJKL} -\textstyle{\frac{1}{2}} p q^2 \sin^2 \alpha \, C_-^{IJKL} - \textstyle{\frac{1}{8}} i p q^2 \sin 2 \alpha \, D_-^{IJKL}
\end{equation}
or, equivalently,
\smallskip
\begin{equation}
 {u^{IJ}}_{KL} (\lambda, \alpha) \,=\,  p^3 \delta^{IJ}_{KL} + \textstyle{\frac{1}{2}} p q^2 \cos^2 \alpha \, C_+^{IJKL} -\textstyle{\frac{1}{2}} p q^2 \sin^2 \alpha \, C_-^{IJKL} + \textstyle{\frac{1}{8}} i p q^2 \sin 2 \alpha \, D_-^{IJKL} \label{u}
\end{equation}
for the complex conjugate.

The derivation of $v^{IJKL}$ is essentially the same as that of ${u_{IJ}}^{KL}$. Starting from equation \eqref{vphi}
\begin{align}
  v^{IJKL} &=\varphi^* \displaystyle{\sum_{n=0}^{\infty}} \frac{(\lambda/2)^{2n+1}}{(2n+1)!} (\varphi \varphi^*)^n \notag \\[2mm]
           &=\textstyle{\frac{1}{2}} \varphi^* \displaystyle{\sum_{n=0}^{\infty}} \frac{\lambda^{2n+1}}{(2n+1)!} \Big\lbrace 1 + \left[3^{2n} - 1 \right] \Theta \Big\rbrace \notag \\[2mm]
           &=\textstyle{\frac{1}{2}} \varphi^* \big\lbrace \sinh \lambda + (\textstyle{\frac{1}{3}} \sinh 3 \lambda - \sinh \lambda) \Theta \big\rbrace, \notag
\end{align}
where the second equality is a direct application of the results derived above.  Substituting for $\varphi^*$ and $\Theta$ using equations \eqref{varphi} and \eqref{deftheta}, respectively, the above expression simplifies to
\begin{align}
 v^{IJKL}\,=\, \textstyle{\frac{1}{48}} (\cos \alpha \, C_+  - i\sin \alpha \, C_-) & \left\lbrace 2 (\sinh 3 \lambda + 9\sinh \lambda) \textstyle{\frac{}{}} \right. \notag \\
 & \left. \hspace{-2mm} + (\sinh 3 \lambda -3 \sinh \lambda)(\cos^2 \alpha \, C_+  -\sin^2 \alpha \, C_-  - \textstyle{\frac{i}{4}} \sin 2 \, \alpha D_-) \right\rbrace
 \end{align}
where we have substituted for $\tilde{\Theta}$ using equation \eqref{tiltheta}.  Expanding out the bracket above and using equations \eqref{C+D+}--\eqref{D-D-}, one can simply show that the above expression reduces to
\begin{align} 
 v^{IJKL} (\lambda,\alpha)\,=\,  q^3 (\cos^3 \alpha + i \sin^3 \alpha) \, \delta^{IJ}_{KL} + & \textstyle{\frac{1}{2}} p^2 q \cos \alpha \,  C_+^{IJKL} -\textstyle{\frac{1}{2}} i p^2 q \sin \alpha \, C_-^{IJKL} \notag \\[5pt]
  -& \textstyle{\frac{1}{8}} q^3 \sin 2 \alpha (\sin \alpha + i \cos \alpha) D_+^{IJKL},  \label{v}
\end{align}
where we have used well-known multiple angle identities for hyperbolic functions and definitions \eqref{defpq}.

The G$_2$ invariant stationary point is given by\footnote{The parametrisation of $\phi_{IJKL}$ used in this paper coincides with that defined in \cite{Warner83, Warner84} by taking $\lambda \rightarrow \textstyle{\frac{1}{2\sqrt{2}}} \lambda$.  Thus, the values of $c$, $s$ and $v$ coincide precisely with those given in \cite{Warner83}.} \cite{Warner83}
\begin{gather}
 c^2 = (p^2 + q^2)^2 = \textstyle{\frac{1}{5}} (3+ 2 \sqrt{3}), \label{cval} \\
 s^2= (2pq)^2 =  \textstyle{\frac{2}{5}} (\sqrt{3}-1), \label{sval} \\
v^2= \cos^2 \alpha = \textstyle{\frac{1}{4}} (3 - \sqrt{3}), \label{vval} 
\end{gather}
where $c = \cosh (2\lambda)$ and $s = \sinh(2\lambda)$.
We also define the following useful combinations
\begin{align}
 b_1(\lambda,\alpha) &= c^3+v^3 s^3, \qquad b_2(\lambda,\alpha) = csv(c+vs), \label{b2} \\
 f_1(\lambda,\alpha) &= p^2 q^2 (p^{2} + q^{2}) \sin \alpha \cos \alpha = \frac{1}{4} s^2 c v \sqrt{1-v^2},  \label{f1}\\
 f_2(\lambda,\alpha) &= p^3 q^3 \sin \alpha \cos^2 \alpha = \frac{1}{8} s^3 v^2 \sqrt{1-v^2},\\
 f_3(\lambda,\alpha) &= p q (p^{2} + q^{2})^2 \sin \alpha = \frac{1}{2} s c^2 \sqrt{1- v^2}. \label{f3}
\end{align}
At the G$_2$ invariant stationary point we have the following simplifying relations
\begin{equation}
 b_1 = 3 b_2, \qquad \qquad f_3 = 4\big( 2 f_1 - f_2\big),   \label{G2simple}
\end{equation}
which will be useful below.

\subsection{Derivation of the 3-form potential}

In this section we derive the potential of the eleven-dimensional G$_2$ invariant solution using the non-linear flux ansatz \eqref{potans}, showing that its field strength coincides with the expression found in \cite{dWNW}.

We identify the expression multiplying the potential in equation \eqref{potans} as
\begin{equation}
8 \Delta^{-1} g^{p q} =  K^{p}{}^{IJ} K^{q}{}^{KL} \left( u^{ij}{}_{IJ} + v^{ij IJ} \right) \left( u_{ij}{}^{KL} + v_{ij KL} \right)
\end{equation}
via the non-linear metric ansatz \eqref{metansatz}. 
Using this, equation \eqref{potans} takes the form
\begin{equation}
A_{mnp} = - \frac{1}{8\sqrt{2}} (i \Delta g_{p q})\, K_{mn}{}^{IJ} K^{q}{}^{KL} \left( u^{ij}{}_{IJ} - v^{ij IJ} \right) \left( u_{ij}{}^{KL} + v_{ij KL} \right). \label{potexp}
\end{equation}

From the metric ansatz \cite{dWNW}
\begin{equation}
 g^{mn} = \frac{\Delta}{6} \left\lbrace \left[ 6 b_1 - b_2 (\xi+3) \right] \go^{mn} + b_2 (21+\xi)  \hat{\xi}^m \hat{\xi}^n \right\rbrace.
\end{equation}
Equivalently,
\begin{equation}
\Delta g_{mn} =  \left\lbrace \frac{6}{6 b_1 - b_2 (\xi +3)} \left (\go_{mn} - \hat{\xi}_m \hat{\xi}_n \right) + \frac{1}{b_1 + 3 b_2} \hat{\xi}_m \hat{\xi}_n \right\rbrace \label{gvonmetans}.
\end{equation}
Substituting the G$_2$ invariant stationary point values given in equations \eqref{cval}--\eqref{vval}
and using (\ref{G2simple})  gives
\begin{equation}
 g_{mn} = 3^{1/6} 10^{1/3} (15-\xi)^{-1/3} \left\lbrace \left(\go_{mn} - \hat{\xi}_m \hat{\xi}_n\right) + \textstyle{\frac{1}{36}}(15 - \xi)\hat{\xi}_m \hat{\xi}_n \right\rbrace,
\end{equation}
which coincides with the metric given in equation \eqref{11dmet}.

In order to simplify the right hand side of equation \eqref{potexp}, recall that $u^{ij}{}_{IJ}$ and $v^{ijIJ}$ are components of E$_7$ matrices. In particular, they satisfy the relations \cite{dWNn8}
\begin{align}
 u^{ij}{}_{IJ} u_{ij}{}^{KL} - v_{ijIJ} v^{ijKL} &= \delta^{KL}_{IJ},\label{e7reln1} \\
 u^{ij}{}_{IJ} v_{ijKL} - v_{ijIJ} u^{ij}{}_{KL} &= 0. \label{e7reln2}
\end{align}
These relations can be explicitly verified for the components of the E$_7$/SU(8) coset element given in equations \eqref{u} and \eqref{v} by using identities \eqref{C+D+}--\eqref{D-D-}. Now, the expression for the E$_7$ matrix components in equation \eqref{potexp}
\begin{align}
 \left( u^{ij}{}_{IJ} - v^{ij IJ} \right) \left( u_{ij}{}^{KL} + v_{ij KL} \right) &= u^{ij}{}_{IJ} u_{ij}{}^{KL} - v^{ij IJ} v_{ij KL} + u^{ij}{}_{IJ} v_{ij KL} - v^{ij IJ} u_{ij}{}^{KL} , \notag\\
&= u^{ij}{}_{IJ} u_{ij}{}^{KL} - (u^{ij}{}_{KL} u_{ij}{}^{IJ} - \delta^{IJ}_{KL}) + u^{ij}{}_{KL} v_{ij IJ} - v^{ij IJ} u_{ij}{}^{KL}. \notag
\end{align}
Recalling that 
\begin{align}
 u_{ij}{}^{IJ} &= \left( u^{ij}{}_{IJ} \right)^*, \qquad v_{ijIJ} = \left( v^{ij IJ} \right)^*, \label{ucomplexcon}
\end{align}
the above expression reduces to
\begin{align}
 \left( u^{ij}{}_{IJ} - v^{ij IJ} \right) \left( u_{ij}{}^{KL} + v_{ij KL} \right)
&= \delta^{IJ}_{KL} + 2 i \textup{Im} \big(u^{ij}{}_{IJ} u_{ij}{}^{KL} - v^{ij IJ} u_{ij}{}^{KL} \big),
\end{align}
where the last term is the imaginary part of the expression in the bracket. 

Using equations \eqref{C+D+}--\eqref{D-D-}, it is straightforward to show that 
\begin{equation}
 \textup{Im} (u^{ij}{}_{IJ} u_{ij}{}^{KL}) = -\frac{1}{4} p^{2} q^{2} (p^{2} + q^{2}) \sin 2 \alpha \,
 D_{-}^{IJKL}. \label{Imuu}
\end{equation}
The expression on the left-hand-side of the above equation is antisymmetric under the exchange of the pair of indices $[IJ]$ and $[KL],$ since from equation \eqref{ucomplexcon} this operation is equivalent to complex conjugation of the expression in the bracket. Therefore, it should come as no surprise that the right-hand-side is given solely in terms of $D_{-}.$ Furthermore,   
\begin{equation}
 \textup{Im} (v^{ij IJ} u_{ij}{}^{KL} ) = 4 p^{3} q^{3} \sin^3 2 \alpha \delta^{IJ}_{KL} -\frac{1}{2} p q (p^{2} + q^{2})^2 \sin \alpha C_{-}^{IJKL} - p^{3} q^{3} \sin \alpha \cos^2 \alpha D_{+}^{IJKL}, \label{Imvu}
\end{equation}
which is indeed symmetric under the exchange of the pairs of indices $[IJ]$ and $[KL]$ as expected from equation \eqref{e7reln2}.

Using equations \eqref{Kcontract}, \eqref{C+}, \eqref{C-} and \eqref{Dpmdef} we derive
\begin{align}
 K_{ab}{}^{IJ} K_{c}{}^{KL} C_{+}^{IJKL} &= - \textstyle{\frac{16}{3}} \delta^{ab}_{cd} \sqrt{(21 + \xi)(3 - \xi)} \hat{\xi}^{d} = -\frac{16}{3} \delta^{ab}_{cd} {\xi}^{d}, \label{KKC+} \\[2mm]
 K_{ab}{}^{IJ} K_{c}{}^{KL} C_{-}^{IJKL} &= 16 S_{abc}, \label{KKC-} \\[2mm]
 K_{ab}{}^{IJ} K_{c}{}^{KL} D_{+}^{IJKL} &= - \textstyle{\frac{8}{3}} (9 + \xi) S_{abc} + 8 (21+ \xi) \hat{\xi}^{d} \hat{\xi}_{[a} S_{bc]d} + \frac{4}{9} {\xi}^{d} \epsilon_{abcdefg} S^{efg}, \label{KKD+} \\[2mm]
 K_{ab}{}^{IJ} K_{c}{}^{KL} D_{-}^{IJKL} &= - \textstyle{\frac{8}{3}} (3-\xi) S_{abc} + 8 (21+ \xi) \hat{\xi}^{d} \hat{\xi}_{[a} S_{bc]d} - \frac{16}{3} (21+ \xi) \hat{\xi}^{d} \hat{\xi}_{c} S_{abd} - \frac{4}{9} {\xi}^{d} \epsilon_{abcdefg} S^{efg}. \label{KKD-}
\end{align}
The first two relations are easily seen to be consistent with equations 
\eqref{defxia} and \eqref{Sdef}. Observe also that the last expression is {\em not}
fully anti-symmetric in the indices $[abc]$.

With the use of the above relations, the expression for the 3-form potential, \eqref{potexp}, reduces to 
\begin{align}
 &A_{mnp} = \frac{1}{18\sqrt{2}} \Delta g_{p q} \go^{qr} \left\{6 \Bigl( (3- \xi) f_1 - 2(9+\xi) f_2 + 6 f_3 \Bigr) \So_{mnr} \notag \right. \\[2pt]
&\hspace{40mm} - 18(21+\xi)(f_1 -2 f_2) \hat{\xi}^{s} \hat{\xi}_{[m} \So_{nr]s} + 12 (21+\xi) f_1 \hat{\xi}^{s} \hat{\xi}_{r} \So_{mns} \notag \\[7pt]
&\hspace{70mm} \left. + \sqrt{(21 + \xi)(3 - \xi)} (f_1 +2 f_2) \hat{\xi}^{s} \etao_{mnrstuv} \So^{tuv} \right\},
\end{align}
where $f_1, f_2, f_3$ are defined in equations \eqref{f1}--\eqref{f3}. The quantities $\hat{\xi}^{m}, \So_{mnp}$ and $\etao_{mnpqrst}$ are constructed with the round $S^7$ vielbein and they are raised/lowered with the round $S^7$ metric, as emphasised earlier.   

Inserting the expression for the metric, equation \eqref{gvonmetans}, found from the non-linear metric ansatz, 
\begin{align}
 &A_{mnp} = \frac{1}{18\sqrt{2}[6 b_1 - b_2 (\xi +3)](b_1 + 3 b_2)} \left\lbrace 6 (b_1 + 3 b_2) \delta_{p}^{q} - (21+ \xi) b_2 \hat{\xi}_p \hat{\xi}^q \right\rbrace \notag \\[6pt]
& \hspace{20mm} \times \left\{ 6 \Bigl( (3- \xi) f_1 - 2(9+\xi) f_2 + 6 f_3 \Bigr) \So_{mnq} \notag \right. \\[3pt] 
&\hspace{40mm} - 18(21+\xi)(f_1 -2 f_2) \hat{\xi}^{r} \hat{\xi}_{[m} \So_{nq]r} + 12 (21+\xi) f_1 \hat{\xi}^{r} \hat{\xi}_{q} \So_{mnr} \notag \\[6pt]
&\hspace{70mm} \left. + \sqrt{(21 + \xi)(3 - \xi)} (f_1 +2 f_2) \hat{\xi}^{r} \etao_{mnqrstu} \So^{stu} \right\},
\end{align}  
where $b_1$ and $b_2$ are defined in equation \eqref{b2}.  Expanding out the terms in the expression above gives
\begin{align}
 &A_{mnp} = \frac{1}{3\sqrt{2}[6 b_1 - b_2 (\xi +3)]} \left\{ 6 \Bigl( (3- \xi) f_1 - 2(9+\xi) f_2 + 6 f_3 \Bigr) \So_{mnp} \notag \right. \\[3pt] 
&\hspace{55mm} - 18(21+\xi)(f_1 -2 f_2) \hat{\xi}^{r} \hat{\xi}_{[m} \So_{np]r} \notag \\[6pt]
&\hspace{65mm} \left. + \sqrt{(21 + \xi)(3 - \xi)} (f_1 +2 f_2) \hat{\xi}^{q} \etao_{mnpqrst} \So^{rst} \right\} \notag \\[5pt]
& \hspace{14mm} + \frac{\sqrt{2}(21+ \xi)}{[6 b_1 - b_2 (\xi +3)](b_1 + 3 b_2)} \left[ 2 b_1 f_1 + b_2 (2 f_1 - 4 f_2 - f_3) \right] \hat{\xi}^{q} \hat{\xi}_{p} \So_{mnq}.
\end{align}  
Let us consider the coefficient of the term that is not totally antisymmetric in the indices $[mnp],$ 
\begin{align*}
 2 b_1 f_1 + b_2 (2 f_1 - 4 f_2 - f_3) &= \frac{1}{2} \sin \alpha\,  vcs^2 \left[ (c^3 + v^3 s^3) - (c + vs) (c^2-vcs+ v^2 s^2) \right]=0,
\end{align*}
where in the first equality we have simply substituted in the definitions of $b_1, b_2, f_1, f_2 $ and $ f_3$ using equations \eqref{b2}--\eqref{f3}. The vanishing of the non-antisymmetric term
even away from the stationary point is expected from the general argument of \cite{dWN13}, 
where it is shown that the 3-form potential as defined by the non-linear flux ansatz is totally antisymmetric by the E$_7$ properties of $u^{ij}{}_{IJ}$ and $v^{ijIJ}.$ 
We now have a totally antisymmetric expression for the 3-form potential
 \begin{align}
 &A_{mnp} = \frac{1}{3\sqrt{2}[6 b_1 - b_2 (\xi +3)]} \left\{ 6 \Bigl( (3- \xi) f_1 - 2(9+\xi) f_2 + 6 f_3 \Bigr) \So_{mnp} \notag \right. \\[3pt] 
&\hspace{55mm} - 18(21+\xi)(f_1 -2 f_2) \hat{\xi}^{r} \hat{\xi}_{[m} \So_{np]r} \notag \\[6pt]
&\hspace{65mm} \left. + \sqrt{(21 + \xi)(3 - \xi)} (f_1 +2 f_2) \hat{\xi}^{q} \etao_{mnpqrst} \So^{rst} \right\}. \label{A3}
\end{align}  
This is only defined up to gauge transformations, hence to make a comparison with the known G$_2$ invariant solution, we calculate its field strength. Using equations \eqref{Dxi}, \eqref{Dxia} and \eqref{DS}, the field strength of the potential above, 
$$F_{mnpq}= 4   \Do_{[m} A_{npq]},$$ is 
\begin{align}
& F_{mnpq} \, = \notag \\[5pt]
& = \frac{4\sqrt{2} m_7}{6 b_1 - b_2(3+ \xi)} \Biggl\{ (f_3- 4 f_2) \etao_{mnpqrst} \So^{rst} \notag \\[7pt]
&\quad -\frac{2(21+\xi)}{3[6 b_1 - b_2(3+ \xi)]} \Bigl[ b_2 (f_1 - f_2 ) (\xi - 27) - 3 b_1 (f_1 - 4 f_2) + 9 b_2 (3 f_1 - 4 f_2) \Bigr] \hat{\xi}_{[m} \etao_{npq]rstu} \hat{\xi}^{r} \So^{stu} \notag \\[7pt]
& \quad -\frac{4 \sqrt{(21+ \xi)(3- \xi)}}{6 b_1 - b_2(3+ \xi)} \Bigl[ b_2 (f_1 - f_2 ) (51 - \xi) + 3 b_1 (f_1 - 4 f_2) - 3 b_2 (17 f_1 -16 f_2 - f_3) \Bigr] \So_{[mnp} \hat{\xi}_{q]} \Biggr\}. \label{Fvonfluxansatz}
\end{align}
Substituting relations (\ref{G2simple}), valid at the G$_2$ stationary point, we get
\begin{align}
 F_{mnpq} = \frac{32\sqrt{2}(f_1-f_2)}{b_2(15-\xi)} m_7 &\left\lbrace \etao_{mnpqrst} \So^{rst} - \frac{(21+\xi)}{12(15-\xi)}(\xi-27 +  \textstyle{\frac{18f_1}{f_1-f_2}})  \hat{\xi}_{[m} \etao_{npq]rstu} \hat{\xi}^r \So^{stu} \right. \notag \\[6pt]
 &\hspace{25mm} +\left.  \frac{\sqrt{(21+\xi)(3-\xi)}}{2(15-\xi)}(\xi-51 +  \textstyle{\frac{18f_1}{f_1-f_2}}) \So_{[mnp} \hat{\xi}_{q]} \right\rbrace.
\end{align}
Using equations \eqref{cval}--\eqref{vval}, the expression above reduces to 
\begin{align}
 F_{mnpq} = \frac{4 \times 3^{-1/4}}{15-\xi} m_7 &\left\lbrace \etao_{mnpqrst} \So^{rst} - \frac{(21+\xi)(\xi-27 + 12 \sqrt{3})}{12(15-\xi)}  \hat{\xi}_{[m} \etao_{npq]rstu} \hat{\xi}^r \So^{stu} \right. \notag \\[6pt]
 &\hspace{25mm} +\left. \sqrt{(21+\xi)(3-\xi)} \frac{(\xi-51 + 12 \sqrt{3})}{2(15-\xi)} \So_{[mnp} \hat{\xi}_{q]} \right\rbrace.
\end{align}
This is in perfect agreement with the flux of the G$_2$ invariant solution \cite{dWNW} given in equation \eqref{g2flux}.  It is remarkable that there is not only an agreement with the general structure, but also the precise coefficients.

\section{The SU(4)$^{-}$ invariant solution} \label{sec:su4}

\subsection{The SU(4)$^{-}$ invariant solution of eleven-dimensional supergravity}

The SU(4)$^{-}$ invariant solution \cite{PW} is a compactification of eleven-dimensional supergravity to a maximally symmetric four-dimensional spacetime with the internal space given by a stretched U(1) fibration over CP$^{3}.$ In \cite{PW}, the solution was expressed in terms of structures on CP$^{3}.$ Here, in order to compare the SU(4)$^{-}$ invariant solution with the result given by the non-linear ans\"atze, we express the SU(4)$^{-}$ invariant solution in terms of geometrical quantities defined on a round $S^{7}.$ 

The antiself-dual SO(8) tensor $Y^{-}_{IJKL}$ satisfying \cite{dWNW}
\begin{align}
 Y^{-}_{IJMN} Y^{-}_{MNKL} &=  8 \delta^{KL}_{IJ} - 8 F^{-}_{[I}{}^{[K} F^{-}_{J]}{}^{L]}, \label{YY} \\[2mm] 
 Y^{-}_{IJKL} Y^{-}_{MNPQ} Y^{-}_{PQKL} &= 16 Y^{-}_{IJPQ} \label{YYY}
\end{align}
preserves SU(4)$^{-}$.  The antisymmetric tensor $F^{-}_{IJ}$ is an almost complex structure, 
\begin{equation}
 F^{-}_{I}{}^{K} F^{-}_{K}{}^{J} = - \delta^{J}_{I}. \label{FF}
\end{equation}

Using the properties of $Y^{-}_{IJKL}$ and $F^{-}_{IJ},$ it is straightforward to show that 
\begin{align}
 Y^{-}_{MIJK} F^{-}_{L}{}^{M} &=   Y^{-}_{M[IJK} F^{-}_{L]}{}^{M}, \label{YFantisym} \\
 Y^{-}_{MIJK} F^{-}_{L}{}^{M} &= - \frac{1}{4!} \epsilon_{IJKLPQRS} Y^{-}_{MPQR} F^{-}_{S}{}^{M}, \label{YFantidual} \\
 F^{-}_{[IJ} F^{-}_{KL]} &= - \frac{1}{4!} \epsilon_{IJKLPQRS} F^{-}_{PQ} F^{-}_{RS}. \label{FFantidual}
\end{align}

The SO(8) objects can be used to define the SO(7) tensors
\begin{align}
K_{a} &= \frac{1}{4} K_{a}^{IJ} F^{-}_{IJ}, \qquad K_{ab} = \frac{1}{4} K_{ab}^{IJ} F_{IJ}, \qquad T_{abc} = \frac{1}{16} K_{[ab}^{IJ} K_{c]}^{KL} Y^{-}_{IJKL}, \label{Tdef}
\end{align}
where $K_{a}^{IJ}$ and $K_{ab}^{IJ}$ have been defined in equation \eqref{Kadef}.  Using the relations given in appendix \ref{app:gamma}, the following identities hold
\begin{align}
 K_{a} K_{a} &= 1, \qquad K_{a} K_{ab} =0, \qquad \hspace{23.5mm} K_{ac} K_{cb} = K_{a} K_{b} - \delta_{ab}, \label{KK1} \\
 K_{a} T_{abc} &= 0, \qquad T^{acd} T_{bcd} = 4 (\delta^{a}_{b} - K^{a} K_{b}), \qquad \epsilon^{abcdefg} K_{d} K_{eh} T_{hfg} = - 6 T^{abc}. \label{TT}
\end{align}

Furthermore, using equation \eqref{spinoreqn}
\begin{align}
 \Do_{a} K_{b} &= - m_7 K_{ab}, \label{DK} \\
\Do_{a} T_{bcd} &= \frac{1}{6} m_7 \epsilon_{abcdefg} T^{efg}. \label{DT}
\end{align}

In terms of the tensors $ K_{a}$ and $T_{abc},$ the internal metric of the SU(4)$^{-}$ invariant solution is given by\footnote{As before, we have fixed the allowed arbitrary scaling \eqref{scaling} in anticipation of the form of the metric given by the non-linear ansatz.}
\begin{equation}
 g_{mn} = 2^{-1/3} (\go_{mn} + \Ko_{m} \Ko_{n}), \label{su4metric}
\end{equation}
where as before $\go_{mn}$ is the round $S^7$ metric and $$\Ko_m = \eo_{m}{}^{a} K_{a}$$ is defined with respect to the siebenbein on the round $S^{7}$. 

Using equations \eqref{KK1} and \eqref{DK}, the Ricci tensor of this metric is given by 
\begin{align}
 R_{mn} = \Ro_{mn} + 2 m_7^2 \go_{mn} - 20 m_7^2 \Ko_{m} \Ko_{n}. \label{riccisu4}
\end{align}
The expression for the Ricci tensor of the round $S^{7}$ metric is given in equation \eqref{riccis7}. 

The internal flux of the SU(4)$^{-}$ invariant solution is 
\begin{equation}
F_{mnpq} = \textstyle{\frac{1}{3}} m_7 \etao_{mnpqrst} \To^{rst}. \label{su4flux}
\end{equation}
To verify that the Einstein equations, \eqref{Einseqn1} and \eqref{Einseqn2}, are satisfied it is useful to note that 
\begin{equation}
 F_{mpqr} F^{npqr} = 48 \times 2^{4/3} m_{7}^{2} (\delta_{m}^{n} + K_{m} K^{n}), \label{fluxsqrd}
\end{equation}
where we have used equations \eqref{TT}. On the left-hand side of the above equation, the indices have been raised with inverse of $g_{mn}$ given in equation \eqref{su4metric}. 

Using the expression for the Ricci tensor in the internal direction, \eqref{riccisu4}, and equation \eqref{fluxsqrd}, it is straightforward to verify that $g_{mn}$ and $F_{mnpq}$ solve the Einstein equations, \eqref{Einseqn1} and \eqref{Einseqn2}, with 
\begin{align}
 m_{4}^2 &= \frac{16}{3} m_7^{2}, \qquad \fr^2 = 32 m_{7}^{2}. \label{fm7}
\end{align}
With the above value for $\fr$, the equation of motion for $F_{mnpq}$, \eqref{maxwell}, is also satisfied.

\subsection{The SU(4)$^{-}$ invariant stationary point of gauged supergravity}

The SU(4)$^{-}$ invariant stationary point of maximal gauged supergravity is obtained for a purely pseudoscalar expectation value $\phi_{IJKL}$ of the form \cite{Warner83}
\begin{equation}
 \phi_{IJKL}=\textstyle{\frac{1}{2}}i\lambda Y^-_{IJKL},
\end{equation}
where $Y^-_{IJKL}$ is an antiself-dual object satisfying the properties presented in equations \eqref{YY}--\eqref{FF}.

Using equation \eqref{YYY}, it is simple to show that for $n>0,$
\begin{equation} \label{YYn}
 {(Y^-Y^-)^n}_{IJKL} = 2^{4(n-1)} {(Y^-Y^-)}_{IJKL},
\end{equation}
where $(Y^- Y^-)_{IJKL}$ denotes a contraction of the form $Y^-_{IJMN} Y^-_{MNKL}$.

As described in section \ref{sec:G2gauged}, it is fairly straightforward to show that
\begin{equation}
{u_{IJ}}^{KL} = \displaystyle{\sum_{n=0}^{\infty}} \frac{(\lambda/2)^{2n}}{(2n)!} {(Y^- Y^-)^n}_{IJKL},
\end{equation}
Using equations \eqref{YYn} and \eqref{YY}, the above expression reduces to
\begin{align}
 {u_{IJ}}^{KL} &=\delta_{IJ}^{KL}+ \displaystyle{\sum_{n=1}^{\infty}} \frac{(\lambda/2)^{2n}}{(2n)!} 2^{4n-1} \left( \delta_{IJ}^{KL} - F^{-}_{[I}{}^{[K} F^{-}_{J]}{}^{L]} \right) \notag \\[5pt]
 &= \delta_{IJ}^{KL}+ \frac{1}{2} \left( \displaystyle{\sum_{n=0}^{\infty}} \frac{(2 \lambda)^{2n}}{(2n)!} -1 \right) \left( \delta_{IJ}^{KL} - F^{-}_{[I}{}^{[K} F^{-}_{J]}{}^{L]} \right) \notag \\[5pt]
 &= \delta_{IJ}^{KL} + \textstyle{\frac{1}{2}} \left( \cosh{2 \lambda} -1 \right) \left( \delta_{IJ}^{KL} - F^{-}_{[I}{}^{[K} F^{-}_{J]}{}^{L]} \right) 
 \end{align}
Defining $c=\cosh(2\lambda)$ as before,  and observing that the expression is real,  
\begin{equation} \label{su4u}
 {u^{IJ}}_{KL} = \textstyle{\frac{1}{2}} (c+1) \delta^{IJ}_{KL} - \textstyle{\frac{1}{2}} (c-1) F^{-}_{[K}{}^{[I} F^{-}_{L]}{}^{J]}.
\end{equation}

Similarly,
\begin{align}
 v^{IJKL} &= - i Y^-_{IJMN} \displaystyle{\sum_{n=0}^{\infty}} \frac{(\lambda/2)^{2n+1}}{(2n+1)!} {(Y^- Y^-)^n}_{MNKL} \notag \\[5pt]
 &= - {\textstyle{\frac{1}{2}}} i Y^-_{IJMN} \left(\lambda+ \frac{1}{32} \displaystyle{\sum_{n=1}^{\infty}} \frac{(2 \lambda)^{2n+1}}{(2n+1)!} {(Y^- Y^-)}_{MNKL}\right) \notag \\[5pt]
 &= - {\textstyle{\frac{1}{2}}} i Y^-_{IJMN} \left(\lambda+ \frac{1}{32} \left[\sinh(2\lambda)-2\lambda\right] {(Y^- Y^-)}_{MNKL}\right) \notag \\[7pt]
  &= - {\textstyle{\frac{1}{4}}} i \sinh(2\lambda) Y^-_{IJKL},
\end{align}
where we have used equation \eqref{YYY} in the final equality in the equation above.  Defining $s=\sinh(2\lambda),$ 
\begin{equation} \label{su4v}
 v^{IJKL}= - {\textstyle{\frac{1}{4}}} i s Y^-_{IJKL}.
\end{equation}
It is simple to verify that the $u$ and $v$ as given in equations \eqref{su4u} and \eqref{su4v} satisfy the E$_7$ relations, equations \eqref{e7reln1} and \eqref{e7reln2}.

From the metric ansatz \cite{dWNW},
\begin{equation}
 \Delta^{-1} g^{mn} = \left\lbrace c^2 \go^{mn} - s^2 \Ko^{m} \Ko^{n} \right\rbrace.
\end{equation}
Equivalently,
\begin{equation} \label{su4metriccs}
\Delta g_{mn} =  c^{-2} \left\lbrace  \go_{mn} + s^2 \Ko_{m} \Ko_{n} \right\rbrace.
\end{equation}
 
The SU(4)$^{-}$ invariant stationary point is given by \cite{Warner83}
\begin{gather} \label{su4cs}
 c^2 = 2, \qquad  s^2= 1. 
\end{gather} 
Substituting these values into the expression above and taking the determinant of the resulting expression gives
\begin{equation}
 \Delta=2^{-2/3}.
\end{equation}
Hence, the metric is of the form
\begin{equation}
 g_{mn} =  2^{-1/3} \left\lbrace  \go_{mn} + \Ko_{m} \Ko_{n} \right\rbrace,
\end{equation}
which agrees with that given in equation \eqref{su4metric}.

Substituting the expression for $u$ and $v$ given in equations \eqref{su4u} and \eqref{su4v}, and the form of the metric given in equation \eqref{su4metriccs} into equation \eqref{potexp}, it is simple to show that
\begin{equation}
 A_{mnp}= - \frac{1}{\sqrt{2}} (s/c)\To_{mnp},
\end{equation}
where we have used the first equation in \eqref{TT}.  Now, using equation \eqref{DT}, the field strength is simply
\begin{equation}
 F_{mnpq}= - \textstyle{\frac{\sqrt{2}}{3}} (s/c) m_7 \etao_{mnpqrst} \To^{rst}.
\end{equation}
Substituting the values of $c$ and $s$ given in equation \eqref{su4cs} gives
\begin{equation}
  F_{mnpq}= - \textstyle{\frac{1}{3}} m_7 \etao_{mnpqrst} \To^{rst}.
\end{equation}
Note that the Einstein equations \eqref{Einseqn1} and \eqref{Einseqn2} and the equation of motion for the flux \eqref{maxwell} are satisfied regardless of the overall sign of the flux.  Thus, again, we have precise agreement with the flux of the SU(4)$^{-}$ solution given in equation \eqref{su4flux}.

\bigskip
\noindent{\bf Acknowledgements:} We would like to thank  Bernard de Wit for discussions.

\newpage
\appendix

\section{SO(7)$^{\pm}$ invariant solutions} \label{app:so7}

For completeness we here reproduce the metric 
and flux of the SO(7)$^{\pm}$ invariant solutions, even though these are simpler than 
the ones  discussed in the text. The relevant solutions can be found in analogy 
with the general metric and flux of the G$_2$ invariant family, given in equations \eqref{gvonmetans} 
and \eqref{Fvonfluxansatz}, and by restricting the scalar fields 
in \eqref{phi} to $\alpha =0$ and $\alpha = \pi/2$, respectively.

The SO(7)$^+$ invariant stationary point of maximally gauged supergravity is given by \cite{Warner83}
\begin{equation}
 c^2=\frac{1}{2} (3/\sqrt{5} +1), \qquad s^2=\frac{1}{2} (3/\sqrt{5} -1), \qquad v=1.
\end{equation}
In particular, these imply that $f_1, f_2, f_3$ as defined in equations \eqref{f1}--\eqref{f3} vanish.  It immediately follows that
\begin{equation}
 F_{mnpq}=0,
\end{equation}
as expected.  The metric is
\begin{equation} \label{so7+met}
 \Delta g_{mn} = \frac{6 \times 5^{1/4}}{9- \xi} \left\lbrace (\go_{mn} - \hat{\xi}_{m} \hat{\xi}_{n}) + \frac{(9-\xi)}{30} \hat{\xi}_{m} \hat{\xi}_{n} \right\rbrace. 
\end{equation}
This is the solution of \cite{dWNso7soln}; see also \cite{dWNW, NP}.  In particular, in reference \cite{dWNW}, the solution is given in the form
\begin{equation}
 \Delta g_{mn} = \frac{30 \gamma^{-1/2}}{9- \xi} \left\lbrace (\go_{mn} - \hat{\xi}_{m} \hat{\xi}_{n}) + \frac{(9-\xi)}{30} \hat{\xi}_{m} \hat{\xi}_{n} \right\rbrace,
\end{equation}
which agrees with metric \eqref{so7+met} for 
$$\gamma=5^{3/2}.$$

Similarly, the SO(7)$^{-}$ stationary point is given by 
\begin{equation}
  c^2=\frac{5}{4}, \qquad s^2=\frac{1}{4}, \qquad v=0.
\end{equation}
Since $v=0 $, $b_2$ as defined in equation \eqref{b2} vanishes and the metric is given by the round $S^{7}$ metric
\begin{equation}
 \Delta g_{mn} = c^{-3} \go_{mn}.
\label{so7-met}
\end{equation}
Moreover the flux for the SO(7)$^{-}$ family is 
\begin{equation}
 F_{mnpq}= \frac{\sqrt{2}}{3} (s/c) m_7 \etao_{mnpqrst} \So^{rst}.
\label{so7-flux}
\end{equation}
This is consistent with the Englert solution \cite{englert}; see also \cite{dWNW, NP}. In reference \cite{dWNW}, the solution is expressed as 
\begin{gather}
  \Delta g_{mn} = \gamma^{-1/2} \go_{mn}, \\
F_{mnpq} = \frac{1}{3 \sqrt{2}} \gamma^{-1/6} m_7 \etao_{mnpqrst} \So^{rst},
\end{gather}
which agree with equations \eqref{so7-met} and \eqref{so7-flux} at the stationary point for 
$$\gamma^{1/3} = 5/4.$$

\section{Useful G$_2$ identities}
\label{app:crelns}

In this appendix, we derive identities relating the contraction of G$_2$ invariants 
$C_{\pm}$ and $D_{\pm}$, adopting the shorthand notation (\ref{AB}) throughout.
In deriving these identities it is useful to observe that viewed as E$_7$ matrices, 
$C_{\pm}$ and $D_+$ are generators of an SU(1,1) subalgebra of E$_7$.  This is 
the unique subalgebra of E$_7$ that commutes with G$_2$ \cite{dWNW}, cf.
\begin{align}
 \sigma^1 \sim \begin{pmatrix}
                0 & C_+ \\ C_+ & 0
               \end{pmatrix}, \qquad
 \sigma^2 \sim \begin{pmatrix}
                0 & - i C_- \\ i C_- & 0
               \end{pmatrix}, \qquad
\sigma^3 \sim \begin{pmatrix}
               D_+ & 0 \\ 0 & - D_+
              \end{pmatrix}.
\end{align}
Thus,
\begin{equation}
 \left[ \begin{pmatrix}
  D_+ & 0 \\ 0 & - D_+
 \end{pmatrix} ,
 \begin{pmatrix}
   0 & C_+ \\ C_+ & 0
 \end{pmatrix} \right] \; \propto \; \begin{pmatrix}
				0 & - i C_- \\ i C_- & 0
				  \end{pmatrix},
\end{equation}
which implies that
\begin{equation}
 (C_+ C_- C_+ + 4 D_+) \propto C_-.
\end{equation}
Consistency with equation \eqref{C+C+} fixes the constant of proportionality:
\begin{equation} \label{C+C-C+}
 C_+ C_- C_+ = -4 (C_-+D_+).
\end{equation}
Similarly,
\begin{equation} \label{C-C+C-}
 C_- C_+ C_- = -4 (C_--D_+).
\end{equation}

Using equations \eqref{C+C+}, \eqref{C-C-}, \eqref{C+C-C+} and \eqref{C-C+C-}, it is straightforward to prove the following identities:
\begin{align}
 C_{+} D_{+} &= 4 C_{-} + 2 D_{-}, \hspace{34mm}  D_{+} C_{+} = 4 C_{-} - 2 D_{-} \label{C+D+} \\
 C_{-} D_{+} &= 4 C_{+} + 2 D_{-}, \hspace{34mm}  D_{+} C_{-} = 4 C_{+} - 2 D_{-}, \\
 C_{+} D_{-} &= 8 C_{-} + 4 D_{+} + 2 D_{-}, \hspace{21.5mm}  D_{-} C_{+} = -8 C_{-} - 4 D_{+} + 2 D_{-}, \\
 C_{-} D_{-} &= - 8 C_{+} + 4 D_{+} - 2 D_{-}, \hspace{18.5mm}  D_{-} C_{-} =  8 C_{+} - 4 D_{+} - 2 D_{-}, \\
 D_{+} D_{+} &= 48  + 8 D_{+}, \hspace{37mm}  D_{-} D_{+} = 16 C_{+} + 16 C_{-} + 4 D_{-}, \label{D-D+} \\
 D_{+} D_{-} &= -16 C_{+} - 16 C_{-} + 4 D_{-}, \hspace{15mm}  D_{-} D_{-} = - 96  - 16 C_{+} + 16 C_{-} - 8 D_{+}. \label{D-D-} 
\end{align}

\section{Seven-dimensional $\Gamma$-matrix identities}
\label{app:gamma}

For the reader's convenience, here we give a list of useful $\Gamma$-matrix identities,
see also the appendices of \cite{cremmerjulia,dWNsu8}. The seven-dimensional, 
Euclidean $8 \times 8$ $\Gamma^{a}$-matrices, where $a$ is a seven-dimensional flat index, satisfy 
\begin{equation}
  \{ \Gamma^{a}, \Gamma^{b} \} = 2 \delta_{ab}.
\end{equation}
The Clifford algebra admits a Majorana representation, which in our conventions corresponds to a purely imaginary representation of the $\Gamma$-matrices. We use a representation in which all $\Gamma$-matrices are hermitian and antisymmetric; or, equivalently, in our representation the charge conjugation matrix is the identity matrix. Moreover,
\begin{equation}
 \Gamma^{abcdefg} = - i \epsilon^{abcdefg} {\bf 1},
\end{equation}
where 
$$ \Gamma^{abcdefg} = \Gamma^{[a} \dots \Gamma^{g]}$$ 
and $\bf 1$ is the $8 \times 8$ identity matrix.
 
The $\Gamma^{a}$ can be regarded as seven out of the eight components of $Spin(8)$ gamma-matrices in a Majorana-Weyl representation. In this way, one can use SO(8) triality to prove the following important relations \cite{cremmerjulia, dWNsu8}
\begin{align}
 \Gamma^{a}_{[AB} \Gamma^{b}_{CD]} &= \frac{1}{24} \epsilon_{ABCDEFGH}  \Gamma^{a}_{EF} \Gamma^{b}_{GH}, \label{gammaabdual} \\
 \Gamma^{a}_{[AB} \Gamma^{a b}_{CD]} &= \frac{1}{24} \epsilon_{ABCDEFGH}  \Gamma^{a}_{EF} \Gamma^{ab}_{GH}, \label{gammaaabdual} \\
 \Gamma^{[a}_{[AB} \Gamma^{bc]}_{CD]} &= - \frac{1}{24} \epsilon_{ABCDEFGH}  \Gamma^{[a}_{EF} \Gamma^{bc]}_{GH}, \label{gammaabcdual} \\
 \Gamma^{[a}_{[AB} \Gamma^{bc]}_{CD]} &= \frac{1}{24} i \epsilon^{abcdefg}  \Gamma^{de}_{[AB} \Gamma^{fg}_{CD]}. \label{gammaabcdual2} 
\end{align}
The uppercase Latin indices are spinor indices and run from 1 to 8.

Further $\Gamma$-matrix identities can be proved using the Fierz identity, which in Euclidean seven-dimensions takes the form
\begin{equation}
 X_{AB} Y_{CD} = \frac{1}{8} \delta_{BC} (X Y)_{AD} - \frac{1}{8} \Gamma^{a}_{BC} (X \Gamma^{a}_{BC} Y)_{AD} +\frac{1}{16} \Gamma^{ab}_{BC} (X \Gamma^{ab}_{BC} Y)_{AD} - \frac{1}{48} \Gamma^{abc}_{BC} (X \Gamma^{abc}_{BC} Y)_{AD},
\end{equation}
where $X$ and $Y$ are arbitrary $8 \times8$ matrices. The identity above is obtained by noting that 
$$\{\delta_{AB}, \Gamma^{a}_{AB}, \Gamma^{ab}_{AB}, \Gamma^{abc}_{AB} \} $$
span the vector space of $8 \times 8$ matrices. 

The Fierz identity can be used to show
\begin{align}
\Gamma^{a}_{AB} \Gamma^{a}_{CD} &= \Gamma^{a}_{[AB} \Gamma^{a}_{CD]} - 2 \delta^{AB}_{CD}, \label{fierzid1} \\
\Gamma^{a}_{AB} \Gamma^{ab}_{CD} + \Gamma^{a}_{CD} \Gamma^{ab}_{AB} &= 2 \Gamma^{a}_{[AB} \Gamma^{ab}_{CD]}, \\
\Gamma^{a}_{AB} \Gamma^{ab}_{CD} - \Gamma^{a}_{CD} \Gamma^{ab}_{AB} &= -4 \left( \delta_{C[A} \Gamma^{b}_{B]D} - \delta_{D[A} \Gamma^{b}_{B]C} \right) , \label{fierzid3} \\
\Gamma^{ab}_{AB} \Gamma^{ab}_{CD} &= 2 \Gamma^{a}_{AB} \Gamma^{a}_{CD} + 16 \delta^{AB}_{CD}, \label{fierzid4} \\
\Gamma^{c(a}_{AB} \Gamma^{b)c}_{CD} &= \frac{6}{5} \Gamma^{c(a}_{[AB} \Gamma^{b)c}_{CD]} - \Gamma^{(a}_{AB} \Gamma^{b)}_{CD} + \frac{1}{5} \delta^{ab} \Gamma^{c}_{AB} \Gamma^{c}_{CD} - \frac{8}{5} \delta^{ab} \delta^{AB}_{CD}, \label{fierzid5} \\
\Gamma^{c[a}_{AB} \Gamma^{b]c}_{CD} &= - \Gamma^{[a}_{AB} \Gamma^{b]}_{CD} - 2 \left( \delta_{C[A} \Gamma^{ab}_{B]D} - \delta_{D[A} \Gamma^{ab}_{B]C} \right), \label{fierzid6} \\
\Gamma^{ca}_{[AB} \Gamma^{bc}_{CD]} &= 5 \Gamma^{a}_{[AB} \Gamma^{b}_{CD]} - \delta^{ab} \Gamma^{c}_{AB} \Gamma^{c}_{CD}. \label{fierzid7} 
\end{align}
Furthermore, it is also useful to note that (see appendix of \cite{dWNsu8})
\begin{align}
 \Gamma^{ab}_{[AB}  \Gamma^{c}_{CD]} \Big|_{-} &=  \Gamma^{[ab}_{[AB}  \Gamma^{c]}_{CD]}, \label{antidualproj1}\\
 \Gamma^{ab}_{[AB}  \Gamma^{cd}_{CD]} \Big|_{-} &=  \Gamma^{[ab}_{[AB}  \Gamma^{cd]}_{CD]}, \label{antidualproj2}
\end{align}
where the vertical bar $|_{-}$ denotes projection to the antiself-dual part. 

\newpage

\bibliography{g2}
\bibliographystyle{utphys}
\end{document}